\documentclass{jpsj3}  
\usepackage{txfonts}

\title{Flux Phase as a Possible Ordered State in $t-J$ Model
 }

\author{\name{Kazuhiro \surname{KUBOKI}}\thanks{E-mail address: kuboki@kobe-u.ac.jp} 
}
\inst{\address{Department of Physics, Kobe University, 
Kobe 657-8501, Japan} 
}

\begin{document}
\maketitle

The competition between several kinds of ordered states is an important problem 
in strongly correlated electron systems. 
In high-temperature cuprate superconductors, antiferromagnetic (AF) and 
superconducting (SC) states can be realized depending on the doping rate ($\delta$), 
and  it has recently been found that, in multilayer cuprate systems, 
they can coexist uniformly in the same CuO$_2$ plane.\cite{Mukuda} 

Whether ordered states other than the AF and SC states exist in cuprates 
is a subtle question concerning the pseudogap in the underdoped region. \cite{pgap1}
Although a state that  has a free energy higher than other states 
cannot occur in principle, there is a possibility that it can 
occur  if the ordered state is suppressed.  
For example, the $d_{x^2-y^2}$-wave  SC order is strongly suppressed 
near a (110) surface so that other states forbidden in the bulk may arise.  
The flux phase in the $t-J$ model can be  such a state.\cite{Affleck,Ogata} 
It is not a stable phase  but is energetically close to the SC state, and thus  
it may occur once the SC order is suppressed. 

The flux phase is a mean-field (MF) solution to the $t-J$ model in which staggered 
circulating currents flow and the flux $\phi$ penetrates the plaquette in a square 
lattice.
Near (away from) half-filling, $\phi=\pm \pi$ 
($\phi \not= \pm \pi$) and the state is called the $\pi$-flux (staggered-flux) phase.  
The circulating current of the flux phase  
will be a staggered current flowing along the (110) surface 
with an amplitude decaying toward the bulk. 
This means that the time-reversal symmetry (${\cal T}$) is broken
locally near the surface, 
and it may explain the results of experiments that suggest ${\cal T}$ 
violation in cuprate superconductors.\cite{TRSB1}
(The $d$-density wave states, which have been introduced in a different context,  
have similar properties.\cite{Chakra})

Zhang examined the stability of the flux phase using the renormalized MF 
theory and found that it is unstable toward the $d$-wave SC instability 
at any finite doping rate.\cite{Zhang} 
Later, Hamada and Yoshioka\cite{Hamada} studied the problem based on 
the slave-boson (SB) MF approximation.\cite{Zou,Lee} 
Their results were essentially the same except that the flux phase may be stable near 
half-filling  when $t/J$ is not large ($t/J \sim 1$). 
Bejas {\it et al.} treated the $t-J$ model with next-nearest-neighbor hopping terms
using 1/N expansion in the leading order. 
In this treatment, the SC and AF states are excluded, and 
it turned out that the flux phase is the leading instability even 
at high doping rates.\cite{Bejas}  

In this short note, we examine the metastability of the flux phase in the $t-J$ model 
with long-range hopping terms (extended $t-J$ model) using the SBMF approximation.
For this purpose, we estimate the bare transition 
temperature of the flux phase, $T_{FL}$, assuming the absence of the SC order.  
The long-range hopping terms are introduced to describe the different shapes of 
Fermi surfaces (FSs) for various cuprate superconductors. 
In this paper, the Bose condensation of holons is assumed, in contrast to Ref. 8,  
in which the same SBMF scheme is employed. 

We consider the extended $t-J$ model whose Hamiltonian is 
given as  
\begin{eqnarray}
\displaystyle 
H = -\sum_{j,\ell,\sigma} 
t_{j\ell} {\tilde c}^\dagger_{j\sigma} {\tilde c}_{\ell\sigma}
 +J\sum_{\langle j,\ell\rangle} {\bf S}_j\cdot {\bf S}_\ell,  
\end{eqnarray}
where the transfer integrals $t_{j\ell}$ are finite for the first-  ($t$), 
second-  ($t'$), and third-nearest-neighbor bonds ($t''$), or zero otherwise.  
$J (>0)$ is  the antiferromagnetic superexchange interaction, 
and $\langle j,\ell \rangle$ denotes nearest-neighbor bonds. 
${\tilde c}_{j\sigma}$ is the electron operator in Fock space without 
double occupancy, and we treat this condition using the SB 
method\cite{Zou,Lee}   
by writing ${\tilde c}_{j\sigma}=b_j^\dagger f_{j\sigma}$ under 
the local constraint $\sum_{\sigma}f_{j\,\sigma}^{\dagger}f_{j\,\sigma} 
+ b_j^{\dagger}b_j = 1$ at every $j$ site. 
Here, $f_{j\sigma}$ ($b_j$) is a fermion (boson) operator  
that carries spin $\sigma$ (charge $e$); the fermions (bosons) are frequently 
referred to as spinons (holons). 

We decouple the above Hamiltonian following the manner in Ref. 8   
by including $t'$ and $t''$. 
Although the bosons are not condensed in purely two-dimensional systems  
at finite temperature ($T$), they are almost condensed at low $T$ 
({\it i.e.}, $T < 3J/16$ 
where the flux phase may occur) and for finite carrier doping ($\delta \gtrsim 0.05$). 
Then we treat them as Bose-condensed.
The bond order parameter (OP) for spinons 
may be a complex number when the flux order occurs, and we denote it as 
$ \sum_\sigma \langle f^\dagger_{j\sigma}f_{j+{\hat x}\sigma} \rangle 
\equiv x_s + i(-1)^{j_x+j_y}y_s$, 
$ \sum_\sigma \langle f^\dagger_{j\sigma}f_{j+{\hat y}\sigma} \rangle 
\equiv x_s -  i(-1)^{j_x+j_y}y_s$,  
where ${\hat x}$ (${\hat y}$) is a unit vector in the $x$ 
($y$)-direction (the lattice constant is taken to be unity) 
with $x_s$ and $y_s$ being real constants. 
A $d$-wave resonating-valence-bond (RVB) OP  on the nearest-neighbor bond
is given as  
$\Delta = \langle f_{j\uparrow}f_{j+{\hat x}\downarrow} - 
f_{j\downarrow}f_{j+{\hat x}\uparrow}\rangle/2
= -\langle f_{j\uparrow}f_{j+{\hat y}\downarrow} - 
f_{j\downarrow}f_{j+{\hat y}\uparrow}\rangle/2$ with $\Delta$ being real.  
Under the assumption of the Bose condensation of holons, $\Delta$ 
is equivalent to the SCOP, and we do not take the AF order
into account 
since we are not interested in the doping region near half-filling 
($\delta \lesssim 0.05$). 
Self-consistency equations for these OPs and the chemical potential 
can be obtained by varying the free energy as in Ref. 8: 
\begin{equation}\begin{array}{rl}
\delta = & \displaystyle \frac{1}{2N}\sum_k 
\Big[F^{(+)}_k D^{(+)}_k  + F^{(-)}_k D^{(-)}_k\Big],  
 \\
 x_s = & \displaystyle \Big(t\delta + \frac{3Jx_s}{8}\Big) \frac{1}{2N} \sum_k
\big(\gamma^{(+)}_k\big)^2 
\frac{\Big[F^{(+)}_k D^{(+)}_k -  F^{(-)}_k D^{(-)}_k\Big]}
{\sqrt{(\varepsilon^r_k)^2+(\varepsilon^i_k)^2}},  \\
y_s = & \displaystyle \frac{3Jy_s}{16N} \sum_k
\big(\gamma^{(-)}_k\big)^2 
\frac{\Big[F^{(+)}_k D^{(+)}_k -  F^{(-)}_k D^{(-)}_k\Big]}
{\sqrt{(\varepsilon^r_k)^2+(\varepsilon^i_k)^2}},  \\
\Delta = & \displaystyle \frac{3J\Delta}{16N}\sum_k \big(\gamma^{(-)}_k\big)^2
\big[F^{(+)}_k + F^{(-)}_k\big].  
\end{array}\end{equation}
where 
\begin{equation}\begin{array}{rl}
\displaystyle D^{(\pm)}_k = & \xi_k 
\pm \sqrt{(\varepsilon^r_k)^2+(\varepsilon^i_k)^2} ,  \\
\displaystyle E^{(\pm)}_k =  &
\sqrt{(D^{(\pm)}_k)^2 +\Delta_k^2}, \\
\displaystyle F^{(\pm)}_k = & 
 {{\rm tanh}(E^{(\pm)}_k/2T)}/{E^{(\pm)}_k},  
\end{array}\end{equation}
with 
$\xi_k =  \displaystyle -\mu - 4t'\delta \cos k_x \cos k_y 
-2t"\delta (\cos 2k_x +\cos 2k_y)$, 
$\varepsilon^r_k =  \displaystyle 
- (2t \delta + 3Jx_s/4) \gamma^{(+)}_k$, 
$\varepsilon^i_k =  \displaystyle 
- 3Jy_s \gamma^{(-)}_k/4$, 
$\Delta_k=  \displaystyle - 3J\Delta  \gamma^{(-)}_k/2$, and 
$\gamma^{(\pm)}_k = \cos k_x \pm \cos k_y$. 
Here, $N$ and $\mu$ are the total number of lattice sites and 
the chemical potential, respectively.
It was shown that the flux phase and SC 
order do not coexist in a uniform system.\cite{Zhang,Hamada}
Thus, we treat them separately and estimate
$T_{FL}$ and the transition temperature of SC, $T_C$, 
by solving Eq. (2) numerically. 

$T_{FL}$ and $T_C$ corresponding to those of 
the LSCO system ($t/J =4, t'/t=-1/6, t''=0$) and   
YBCO system  ($t/J = 4, t'/t=-1/6, t''/t=1/5$) as well as those in the case of 
the simple $t-J$ model ($t/J =4, t'=t''=0$)  are shown in Fig. 1. 
Here, we used a simplified parametrization of $t'$ and $t''$ 
to reproduce the FS,\cite{tana} 
neglecting  bilayer splitting in YBCO.
It is seen that $T_{FL}$ is lower than $T_C$ for the same system 
at any finite $\delta$, 
whereas in Ref. 8 $T_{FL}$ can be higher than  $T_C$ near half-filling.  
The reason for the difference is as follows. 
In this study, the Bose condensation of holons is assumed, since 
we are interested in the region where this assumption is 
justified. 
However, near half-filling, holons are not close to condensation.
Then the flux phase is favored, because the bond order parameters of 
holons can also develop finite imaginary parts. 
At high doping rates, $T_{FL}$ shows a reentrant behavior at a low $T$.
This is because the nesting condition for the FS is changed at a 
high $\delta$ and then the incommensurate flux order, 
which is not taken into account 
in the present work, will be more favorable. 
 
As seen in Fig. 1, 
$T_{FL}$ for the LSCO system remains finite in a rather large doping range 
extending to $\delta \sim 0.146$, much larger than that in the case of $t'=t''=0$. 
This result is consistent with that in  Ref. 11.
In  contrast,  $T_{FL}$ for the YBCO system is limited to a narrower region. 
These differences arise owing to the shape of FS. 
Flux phases are characterized by the imaginary part of 
the bond OP, $y_s$. From Eq. (2), we see that the expression of $y_s$ has 
the form factor $(\gamma^{(-)}_k)^2$ so that, if $|\gamma^{(-)}_k|$ is large 
near the FS, the flux phase should be favored. 
The FS of the LSCO (YBCO) system is favorable (unfavorable) in this sense.\cite{tana} 
%%%%%{FIG.}%%%%%%
\begin{figure}%[htb]
\begin{center}
\includegraphics[width=7.5cm,clip]{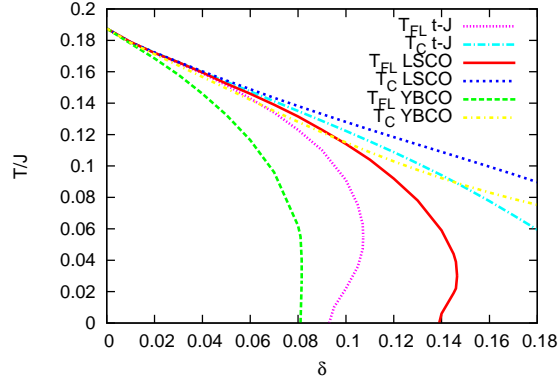}
\caption{(Color online) Bare transition temperature of the flux phase, $T_{FL}$,
and superconductivity, $T_{C}$, for LSCO-type ($t/J=4, t'/t=-1/6, t''=0$), 
YBCO-type ($t/J=4, t'/t=-1/6, t''/t=1/5$), and simple $t-J$ 
($t/J=4, t'=t''=0$) model. 
	 }
\end{center}
\end{figure}

The above results indicate that the flux phase may appear if the SC order is suppressed, 
{\it e.g.}, near  (110) surfaces, 
and it would lead to time-reversal symmetry breaking. 
We might expect ${\cal T}$ violation for surfaces with other types of orientations.
In real systems, 
surfaces may not be so smooth, and so there may be small domains where 
the angle of crystal axes is 45$^\circ$ to the surface.  
In this case, the flux phase may appear leading to a local ${\cal T}$ violation 
in these domains.  

Experimentally, ${\cal T}$ violation is observed not only below but also 
above  $T_C$ in the underdoped region.\cite{TRSB1} 
In order to understand this, the dynamics of holons should be included and 
the effect of fluctuations around the SBMF solution ($U(1)$ gauge fluctuations) 
must be examined. This problem will be studied separately in the future. 

In summary, we have determined the bare transition temperature of the flux phase,   
$T_{FL}$,  in the extended $t-J$ model assuming that the superconducting order 
is absent. It is found that $T_{FL}$ may be finite for a rather large doping range   
if the shape of the Fermi surface is favorable to this state.
In order to clarify whether the flux phase actually appears as a surface state, 
we must examine the spatial variations in the order parameters. 
Calculations based on the Bogoliubov de Gennes method are now under way 
and we will report them separately. 

This work was supported by JSPS KAKENHI Grant Number 24540392.

%----------------------------------------------------------------------------------------------------------

\end{document}